\documentclass[preprint,aps,pre,letterpaper]{revtex4-1}

\usepackage{graphicx}
\usepackage{amsmath}
\usepackage{amsfonts}
\usepackage{amssymb}
\usepackage{epstopdf}

\newcommand{\beq}{\begin{equation}}
\newcommand{\eeq}{\end{equation}}
\newcommand{\bea}{\begin{eqnarray}}
\newcommand{\eea}{\end{eqnarray}}

\begin{document}

\title{Grand-canonical Monte-Carlo simulation of solutions of salt mixtures: theory and implementation}

\author{Toan T. Nguyen$^{1,2,3}$}

\affiliation{
$^1$VNU Key Laboratory "Multiscale Simulation of Complex Systems''
and \\
$^2$Faculty of Physics, VNU University of Science, Vietnam National University, 
334 Nguyen Trai Street
Thanh Xuan, Hanoi, Vietnam.\\
$^3$School of Physics, Georgia Institute of Technology, 
837 State Street, Atlanta, Georgia 30332-0430, USA.\\
}

\begin{abstract}
A Grand-canonical Monte-Carlo simulation method extended to simulate a mixture of salts 
is presented. Due to charge neutrality requirement of electrolyte solutions, 
ions must be added to or removed from the system in groups. 
This leads to some complications compared to regular Grand Canonical simulation. 
Here, a recipe for simulation of electrolyte solution of salt mixture is presented. 
It is then implemented to simulate solution of 1:1, 2:1 and 2:2 salts or their mixtures 
at different concentrations using the primitive ion model. The osmotic pressures of 
the electrolyte solutions are calculated and shown to depend linearly on the salt 
concentrations within the concentration range simulated. We also show that at the same 
concentration of divalent anions, the presence of divalent cations make it easier 
to insert monovalent cations into the system. This can explain some quantitative
differences observed in experiments of the MgCl$_2$ salt mixture and MgSO$_4$ salt mixture.
\end{abstract}



\maketitle

\section{Introduction}

Computer simulation is an integral part of many areas of modern interdisciplinary research 
in physics, chemistry, biology and material science \cite{AllenAndTildesley}. This is especially true for 
computer simulation of biological systems in medicine such as drug design and bioinspired 
novel materials and nanotechnology for medicine \cite{coveney2014computational}. 
For such systems, molecular dynamics 
has been an important computational tool to understand physical characteristics of 
ligand-receptor binding processes, and to predict structural, dynamical and thermodynamic
properties of biological molecules. However, although computing hardware has been 
steadily improved over the year, the large amount of atoms (correspondingly, 
the number of degrees of freedoms) in such system has rendered traditional molecular
dynamics simulation to limited applications within few hundred nanoseconds and tens 
of nanometer scales. This computing requirement is even more demanding and challenging 
when the physics phenomenon involved require quantum mechanical simulation. 
To overcome such limitation and to bridge to larger time and spatial scales, 
multi-scale simulation strategies have been an active research. 
Among them, methods of hybrid Quantum mechanics/Molecular mechanics or 
Coarse-grained/Molecular Mechanics simulation, or Adaptive resolution 
simulation have been proposed with limited success\cite{praprotnik2013multiscale,
AdResHPRL,PaoloCGMM,noid2013perspective,brunk2015mixed}.

The general idea behind multiscale simulation is to focus in molecular details
to only a small, well-defined region (MM region) of interest while the rest of the system
can be simulated at a coarser scale, making the computation
more efficient. The bridging of macromolecules (such as protein or DNA)
between two different scaled regions can be handled adequately in such
hybrid simulation with suitable choice of coarse-grained model
such as the G{\=o} model \cite{Gomodel,hoang2000molecular} for
protein or similar coarse-grained model for DNA \cite{DNACGSchulten}. 
This multiscale strategy also helps to avoid unnecessary bias due to 
potentially wrong orientations of the side chains far from the binding site.
However, the simulation of mobile molecules, especially mobile
ions, into and out of the MM region is still an open question which
is not trivial to handle in a molecular dynamic simulation. In fact,
one usually forbids the mobile ions to move in and out of the MM region
in such simulation. One idea to overcome this is to look beyond molecular dynamics.
Specifically, in addition to molecular dynamics simulation, 
one could try to implement a Monte-Carlo simulation 
in the Grand canonical ensemble. In such simulation, mobile
ions could be inserted and removed from the MM region in such a way
that their chemical potentials are fixed, and controlled by coupling to a particle
reservoir with the correct concentration. 
This is actually desirable because
all biological systems function in equilibrium with water solutions at
given pH and salinity. Of course, developing and implementing such scheme 
for application in computational biomedicine or pharmaceutical nanotechnology
require large amount of time and resources and it is a very active research area.

In this paper, as a first step in such direction, we present a Grand canonical
Monte-Carlo (GCMC) simulation of electrolyte solutions for different salinity expanding upon a preliminary study of single salt electrolyte solution \cite{CohenGCMC}. 
We generalize them to different salt mixtures and present a detailed 
implementation of 1:1, 2:1, and 2:2 salt solutions and their mixture 
(such as both divalent and monovalent anions and cations are present).
The fugacities of various type of salts in different solution and their 
osmotic pressure are studied. Additionally, the effect of additional salts
on the fugacity of a given salt is studied. Our results show that, 
in asymmetric salt mixture, divalent cations can make it easier 
to insert monovalent cations into the solution.

The Grand-Canonical Monte-Carlo method was developed and 
used in several recent papers in our group to study the 
condensation of DNA inside bacteriophages in the presence 
of mixture of different salts, MgSO$_4$, MgCl$_2$, NaCl
\cite{NguyenPRL2010,NguyenJBP2013,NguyenJCP2016,ToanJBP2017}.
However, detail of the method was never presented, only
the simulation results of DNA system were shown. In this paper, the
methodology and implementation of this GCMC method generalized
for salt mixture is presented systematically and 
in detail. This allows for extension to any systems, not just DNA systems,
and for potential integration in various multiscale simulation schemes.

The paper is organized as follows. In Sec. \ref{sec:GCMC}, the Grand-canonical 
Monte-Carlo is formulated to simulate a system of salts mixture.
In Sec. \ref{sec:GCMC2}, the detail implementation of this method
for various salts and salt mixtures are presented. Result for the
fugacities and osmotic pressure are reported and discussed.
We conclude in Sec. \ref{sec:conclusion}. 

\section{Grand canonical Monte$-$Carlo Simulation
of electrolyte solutions : theoretical aspect}

\label{sec:GCMC}
In a Grand Canonical Monte-Carlo (GCMC) simulation,
the number of ions is not constant during the simulation. Instead
their chemical potentials are fixed. 
%
To show how this is done, let us consider a state $i$ of the system that is characterized by
the locations of $N_{iZ+}$ multivalent counterions, 
$N_{i+}$ monovalent counterions, $N_{iZ-}$ multivalent counterions, 
$N_{i-}$ coions (in the next section where we focus on
divalent counterions and coions, $Z=2$).
In the grand canonical ensemble of unlabeled particles, 
the probability of such state is given by
\bea
\pi_i &=& \frac{1}{\cal Z} 
  \frac{1}{\Lambda_{Z+}^{3N_{iZ+}} \Lambda_+^{3N_{i+}} \Lambda_{Z-}^{3N_{iZ-}} \Lambda_-^{3N_{i-}}}
  \nonumber\\
&&\quad \exp \left[\beta(\mu_{Z+}N_{iZ+}+\mu_+N_{i+}
	+\mu_-N_{i-}+\mu_{Z-}N_{iZ-}-U_i\right]
\label{GCMCpii}
\eea
Here, $\cal Z$ is the grand canonical partition function,
$\beta=1/k_BT$, $\Lambda_{x}\equiv h/\sqrt{2\pi m_{x} k_BT}$
are the thermal wavelength of the corresponding ion type (here $x$
are either $Z+,\ Z-,\ -$ or $+$), $U_i$ is the interaction energy of the state $i$,
and $\mu_{x}$ are the corresponding chemical potential of the
ions.

In a standard Monte Carlo simulation, one would like to
generate a Markov chain of system states $i$ with
a limiting probability distribution proportional
to $\pi_i$. To do this, given a state $i$, one tries to move to state $j$
with probability $p_{ij}$. A sufficient
condition for the Markov chain to have the correct limiting
distribution is:
\begin{equation}
\frac{p_{ij}}{p_{ji}} = \frac{\pi_i}{\pi_j}
\end{equation}
As usual, at each step of the chain, a ``trial" move to change
the system from state $i$ to state $j$ is attempted with probability $q_{ij}$
and is accepted with probability $f_{ij}$. Clearly,
\begin{equation}
p_{ij} = q_{ij} f_{ij}
\end{equation}
It is convenient to regard the simulation box as consisting of
$V$ discrete sites ($V$ is very large). Then for a trial move
where $\nu_{\alpha}$ particles of species $\alpha$ are added
to the system:
\begin{equation}
q_{ij} = \frac{1} {V^{\nu_\alpha} \nu_\alpha!}
\end{equation}
Conversely, if $\nu_{\alpha}$ particles of species $\alpha$ are
removed from the system:
\begin{equation}
q_{ij} = \frac{ (N_\alpha - \nu_\alpha)! } {N_\alpha! \nu_\alpha!}
\label{GCMCqijdelete}
\end{equation}

Putting everything together, equations 
(\ref{GCMCpii})$-$(\ref{GCMCqijdelete}) give
us a recipe to calculate the Metropolis acceptance
probability of a particle insertion/deletion move in
GCMC simulation. For example, if in a transition
from state $i$ to state $j$, a multivalent salt molecule
(one $Z-$ion and $Z$ coions) is added to the system,
the Metropolis probability of acceptance of such
move can be chosen as:
\begin{equation}
f_M = \min\{1, ~ f_{ij}/f_{ji} \}
\end{equation}
where
\begin{equation}
\frac{f_{ij}}{f_{ji}} = \frac{B_{Z:1} }
  {(N_{iZ+}+1)(N_{i-}+1)...(N_{i-}+Z)}
   \exp[ \beta (U_i-U_j) ],
\end{equation}
with
\begin{equation}
\label{eq:BZ1}
B_{Z:1} = \exp(\beta\mu_{Z:1}) \frac{V^{Z+1}}{\Lambda_{Z+}^3 \Lambda_-^{3Z}},
\end{equation}
and $\mu_{Z:1} = \mu_{Z+} + Z\mu_-$
is the combined chemical potential of a $Z:1$ salt molecule.
On the other hand, if a multivalent salt molecule 
(one $Z-$ion and $Z$ coions) is removed from the system, we have:
\begin{equation}
\frac{f_{ij}}{f_{ji}} = \frac{N_{iZ+}N_{i-}...(N_{i-}-Z+1)}
  {B_{Z:1}} \exp[ \beta (U_i-U_j) ],
\end{equation}

Similar expressions are easily obtained from addition/removal of
$Z:Z$ salt. For addition,
\begin{equation}
\frac{f_{ij}}{f_{ji}} = \frac{B_{Z:Z} }
  {(N_{iZ+}+1)(N_{iZ-}+1)}
   \exp[ \beta (U_i-U_j) ],
\end{equation}
and for removal,
\begin{equation}
\frac{f_{ij}}{f_{ji}} = \frac{N_{iZ+}N_{iZ-}}
  {B_{Z:Z}} \exp[ \beta (U_i-U_j) ],
\end{equation}
where
\bea
\label{eq:BZZ}
B_{Z:Z} &=& \exp(\beta\mu_{Z:Z}) \frac{V^2}{\Lambda_{Z+}^3 \Lambda_{Z-}^3}.
\eea
and $\mu_{Z:Z} = \mu_{Z+} + \mu_{Z-}$ is the combined chemical potential of $Z:Z$ salt molecule.

For the addition of monovalent $1:1$ salt to the system
\begin{equation}
\frac{f_{ij}}{f_{ji}} = \frac{B_{1:1} }
  {(N_{i+}+1)(N_{i-}+1)}
   \exp[ \beta (U_i-U_j) ],
\end{equation}
and for removal of $1:1$ salt,
\begin{equation}
\frac{f_{ij}}{f_{ji}} = \frac{N_{i+}N_{i-}}
  {B_{1:1}} \exp[ \beta (U_i-U_j) ],
\end{equation}
where
\bea
\label{eq:B11}
B_{1:1} &=& \exp(\beta\mu_{1:1}) \frac{V^2}{\Lambda_+^3 \Lambda_-^3},
\eea
and $\mu_{1:1} = \mu_+ + \mu_-$ is the combined chemical potential of $1:1$ salt molecule.

Because we are trying to simulate a mixture of salts,
to improve the system relaxation and to
improve the sampling of the system's phase space,
in addition to adding/removing of salt molecules, one could add mixed Monte Carlo
moves by both removing and adding ions of different species in a single step, so long as to maintain the charge neutrality. Most simple MC moves that one can add in the simulation is following:
(a) one $Z+$ multivalent anion is added to (or removed from) the system and $Z$ monovalent anions
are removed from (or added to) the system; 
(b) one $Z-$ multivalent cation is added to (or removed from) the system and $Z$ monovalent cations
are removed from (or added to) the system.
The acceptance probabilities of such moves are also easily calculated in the same manner.
For example, if one $Z+$ anions is added to the system and $Z$ monovalent anions
are removed the system, the Metropolis acceptance ratio is:
\begin{equation}
\frac{f_{ij}}{f_{ji}} = \frac{B_{1:1}^Z N_{i+}...(N_{i+}-Z+1)}
  {B_{Z:1}(N_{iZ+}+1)}
   \exp[ \beta (U_i-U_j) ],
\end{equation}
Vice versa, for a ``trial" Monte-Carlo move where one $Z-$ cation is
removed from the system and $Z$ monovalent cation
are added to the system, the Metropolis acceptance ratio is:
\begin{equation}
\frac{f_{ij}}{f_{ji}} = \frac{B_{Z:1} N_{iZ+}}
  {B_{1:1}^Z (N_{i+}+1)...(N_{i+}+Z)}
   \exp[ \beta (U_i-U_j) ].
\end{equation}

Note that, in all the Metropolis acceptance above, one only needs
 3 chemical potentials for the combined salts $\mu_{Z:1}$, $\mu_{Z:Z}$, and $\mu_{1:1}$,
instead of four chemical potentials for individual
ion species, $\mu_x$. This is because the system must maintain charge neutrality in
all addition/deletion moves so all four chemical potentials are not independent.
In our actual implementation,
the scaled fugacities $B_{Z:1}$, $B_{Z:Z}$ and $B_{1:1}$, are used 
instead of the chemical potentials. 

Beside particle addition/deletion moves, one also
tries standard particle translation moves. They are carried
out exactly like in the case of a canonical Monte-Carlo simulation.
In a ``trial" move from state $i$ to state $j$,
an ion is chosen at random and is moved to a random
position in a volume element surrounding its original
position. The standard Metropolis probability
is used for the acceptance of such ``trial" move:
\begin{equation}
f_M = \min \{1, ~ \exp[\beta(U_i-U_j)]\}.
\end{equation}

\section{Grand canonical Monte$-$Carlo Simulation
of electrolyte solution\label{sec:GCMC2}: implementation for different salt mixtures}
\label{sec:simulationDetail}

In this section, the application of the grand$-$canonical Monte$-$Carlo simulation 
detailed in previous section to simulate a bulk concentration of electrolyte mixtures
is presented. We will focus on the cases of 1:1, 2:1 and 2:2 salt solution
and their mixtures. For simplicity, all ions have radius of $\sigma_x=2$\AA.
The primitive ion model is used. The aquaous solution
is modeled implicitly as a continous medium
with dielectric constant, $\varepsilon$.
The interaction between two ions $\alpha$ and $\beta$ with 
radii $\sigma_{\alpha,\beta}$ and charges 
$Q_{\alpha,\beta}$ is given by
\begin{equation}
U = \left\{
  \begin{array}{l l}
    \dfrac{Q_\alpha Q_\beta} {\varepsilon r_{\alpha\beta}}& \quad 
\mbox{if $r_{\alpha\beta} > \sigma_\alpha + \sigma_\beta$}\\
    \infty & \quad \mbox{if $r_{\alpha\beta} < \sigma_\alpha + \sigma_\beta$}\\ 
  \end{array}
    \right.
\end{equation}
where $r_{\alpha\beta}=|\mathbf{r}_\alpha-\mathbf{r}_\beta|$ is the distance
between the ions. 

The simulation is carried out using the periodic boundary condition.
The long-range electrostatic interactions
between charges in neighboring cells are treated using the standard
Ewald summation method \cite{EwaldSum}. 

To be able to calculate the pressure of the system,
the Expanded Ensemble method \cite{Nordenskiold95, NordenskioldJCP86} is implemented. 
This method allows us to
calculate the difference of the system free energies at
different volumes by sampling these volumes simultaneously 
in a simulation run. By sampling two nearly equal volumes, 
$V$ and $V+\Delta V$, and calculate the free energy difference 
$\Delta \Omega$, we can calculate the total pressure of the system:
\begin{equation}
P(T,V,\{\mu_x\}) = - \left.
      \frac{\partial \Omega(T,V,\{\mu_x\})}{\partial V}
    \right|_{T,\{\mu_x\}} 
    \simeq -\frac{\Delta \Omega}{\Delta V}
\label{eq:pressure}
\end{equation}
The volume derivative are taken at constant values of
all four chemical potentials, $\{\mu_x\} \equiv 
\{\mu_{Z+}, \mu_{Z-}, \mu_{+}, \mu_{-}\}$.

For each simulation run, 100 million MC moves are 
carried out depending on the average number of ions in the
system. To ensure thermalization, 10 million initial moves 
are discarded before doing statistical analysis of 
the result of the simulation. All simulations are done using the physics simulation library
SimEngine develop by one of the author (TTN).
This library use OpenCL and OpenMP extensions of the C programming
language to distribute computational workloads on multi-core
CPU and GPGPU to speed up the simulation time. Both molecular dynamics
and Monte-Carlo simulation methods are supported. In this paper
the Monte-Carlo module of the library is used.

\subsection{Finite size effect}

\begin{figure}[ht]
\centering
\resizebox{8.5cm}{!}{\includegraphics{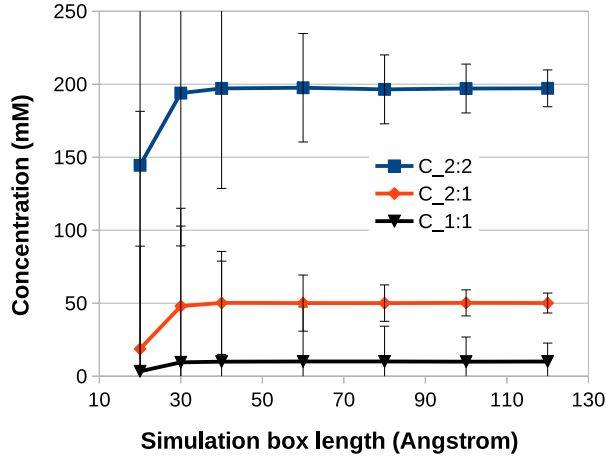}}
\caption{The concentrations of various component salt
in a mixture of three different salts: 2:2, 2:1 and 1:1 salts.
The chemical potentials of salt molecules are fixed. The size of the simulation
box varies from 120\AA\ down to 20\AA. Size dependent effect is only observed
for very small simulation volume such that, on average, there is
less than one salt particle in the volume.
}
\label{fig:Ldependence}
\end{figure}

The first question one asks is the limit of application of this GCMC method. For large
system where the fluctuation in the particle number is fractionally small, 
the simulation result should give the same statistical property of
canonical system. However, for small system where the particle
number fluctuation is large, one might question of validity of the proposed
method. To investigate this finite size effect, we simulate a  
salt mixture at the same chemical potentials (resulting in the same expected
salt concentrations), but with different volume dimensions. Specifically, 
the scaled fugacities are 
$B_{2:2}/V^2=2.05\times 10^{-10}\mbox{\AA}^{-2}$,
$B_{2:1}/V^3=1.14\times 10^{-14}\mbox{\AA}^{-3}$ and
$B_{1:1}/V^2=5.50\times 10^{-10}\mbox{\AA}^{-2}$.
These values are chosen so that, the solution contains a mixture of three
different salts, 2:2 salts, 2:1 salt and 1:1 salt
with the corresponding desired concentrations of
approximately 200mM, 10mM and 50mM (note that, the scaled fugacities $B_{m:n}$, 
eqs. (\ref{eq:BZ1}), (\ref{eq:BZZ}) and (\ref{eq:B11})
must be scaled appropriately for different volumes).
The simulation box is a cubic box with side length varying from 20\AA\ to
120\AA, corresponds to the average number of
particle of divalent anions from 0.7 to about 215.6.
In Figure \ref{fig:Ldependence}, the resultant concentrations at a given
chemical potential is plotted as function of simulation
box lengths. Similarly,
Table \ref{table:Ldependence} shows the numerical values obtained
from our simulation for the averaged concentrations, particle numbers and osmotic
pressures as function of the simulation box lengths.
\begin{table*}[ht]
\centering
\begin{tabular}{c||c|c|c|c|c|c}
\hline
Box length (\AA) & $c_{2:2}$(mM) & $c_{2:1}$(mM) & $c_{1:1}$(mM) & 
	$N_{2+}$ & 	$N_{1+}$ & $P_b$ (atm)\\
\hline
120 & $197.2\pm 12.6$ & $10.0\pm42.7$ & $50.1\pm 6.8$ & 
	$215.60\pm 13.16$ & $52.11\pm 7.11$ & $8.66\pm 0.20$ \\
100 & $197.0\pm 16.7$ & $9.9\pm 16.9$ & $50.2\pm 8.9$ & 
	$124.64\pm 10.16$ & $30.21\pm 5.37$ & $8.59\pm 0.10$ \\
80 &  $196.4\pm  23.6$ & $10.1\pm 24.1$ & $50.0\pm 12.5$ & 
	$63.67\pm 7.44$ & $15.43\pm 3.84$ &$8.73\pm 0.15$\\
60 &  $197.6\pm  37.2$ & $10.1\pm 15.8$ & $50.0\pm 19.2$ & 
	$27.00\pm 4.86$ & $6.51\pm 2.50$ & $8.55\pm 0.16$\\
40 &  $197.1\pm  68.5$ & $9.9\pm 68.9$ & $50.2\pm 35.3$ & 
	$7.98\pm 2.65$ & $1.93\pm 1.36$ & $8.76\pm 0.04$\\
30 &  $193.9\pm 104.7$ & $9.5\pm 105.5$ & $48.0\pm 54.9$ & 
	$3.31\pm 1.72$ & $0.78\pm 0.89$ & $8.53\pm 0.18 $\\
20 &  $144.5\pm 175.6$ & $3.3\pm 178.1$ & $18.7\pm 70.4$ & 
	$0.71\pm 0.86$ & $0.09\pm 0.34$ & $3.84\pm 0.10$ \\
\hline
\end{tabular}
\caption{The result salt concentrations and osmotic pressure of
the solution as function of the length of the cubic simulation box.
The chemical potential are fixed to have the desired mixture of
concentrations of 200mM, 10mM and 50mM for 2:2 salt,
2:1 salt and 1:1 salt correspondingly.}
\label{table:Ldependence}
\end{table*}
We can see that, within the uncertainty of the results, 
all concentrations are independent of simulation box length
down to a very small box length. One only sees the finite
size effect at simulation box length of about 30\AA\ or smaller. 
At these small volumes, the average number of salt molecules 
in the simulation box is even smaller than one for some types 
(such as the number of $+1$ ions as shown in 
Table. \ref{table:Ldependence} ).
This suggests that as long as 
the simulation box are large enough to have a few
ions in it on average, the grandcanonical Monte$-$Carlo 
method presented is reliable. For a given desired concentration, 
the chemical potential of the salts 
are independent on the sizes and shapes of the simulation box.

It should be mentioned here the obvious effect of 
reducing the simulation box size is the increase in the relative fluctuation in 
concentrations. This is inline with statistical theory which says
that the particle number fluctuation increases as $\sqrt{N}$ 
with the number of  particle, $N$. The columns 5 and 6 of 
Table \ref{table:Ldependence} clearly show this quantitative
trend. Because of this, the number fluctuation increases relatively
as $1/\sqrt{N}$ as $N$ decreases. The error bar in Fig. 
\ref{fig:Ldependence} becomes very large at small simulation box size.
Impressively, colume 5 and 6 of Table \ref{table:Ldependence} show
that the $\sqrt{N}$ estimate for fluctuation in the number of particles 
works even for the case the average number of ions is
smaller than one.

In the rest of this paper, 
the simulation box volume is fixed $V=2.650\times 10^3\mbox{ nm}^3$,
corresponding to a box length of 138.4\AA, more than enough
to eliminate possible finite size effects even at some small
salt concentrations simulated.

\subsection{Single salt solution}
Let us present the result of our GCMC simulations for solution
containing a single type of salt, either 1:1, 2:1 or 2:2 salt. 
Some concentrations simulated are already performed
independently by the authors of Ref. \onlinecite{CohenGCMC}. 
For these concentrations, our results agree with their results.
Thus, this section also serves as a check on the correctness 
of our code implementation. 

Tables \ref{table:mu_single11}, \ref{table:mu_single21}, and \ref{table:mu_single22}
show the scaled fugacity $B$ and the resultant
averaged concentration of
the solution obtained from simulation using these parameters.
Three different salts, $1:1$ salt, $2:1$ salt and $2:2$ salt
are listed. 
Standard deviations in the
concentration are about 10\% in our simulation. This relative error is
in line with those of previous GCMC simulations 
of Ref. \onlinecite{CohenGCMC}.

\begin{figure}[ht]
\centering
\resizebox{8.5cm}{!}{\includegraphics{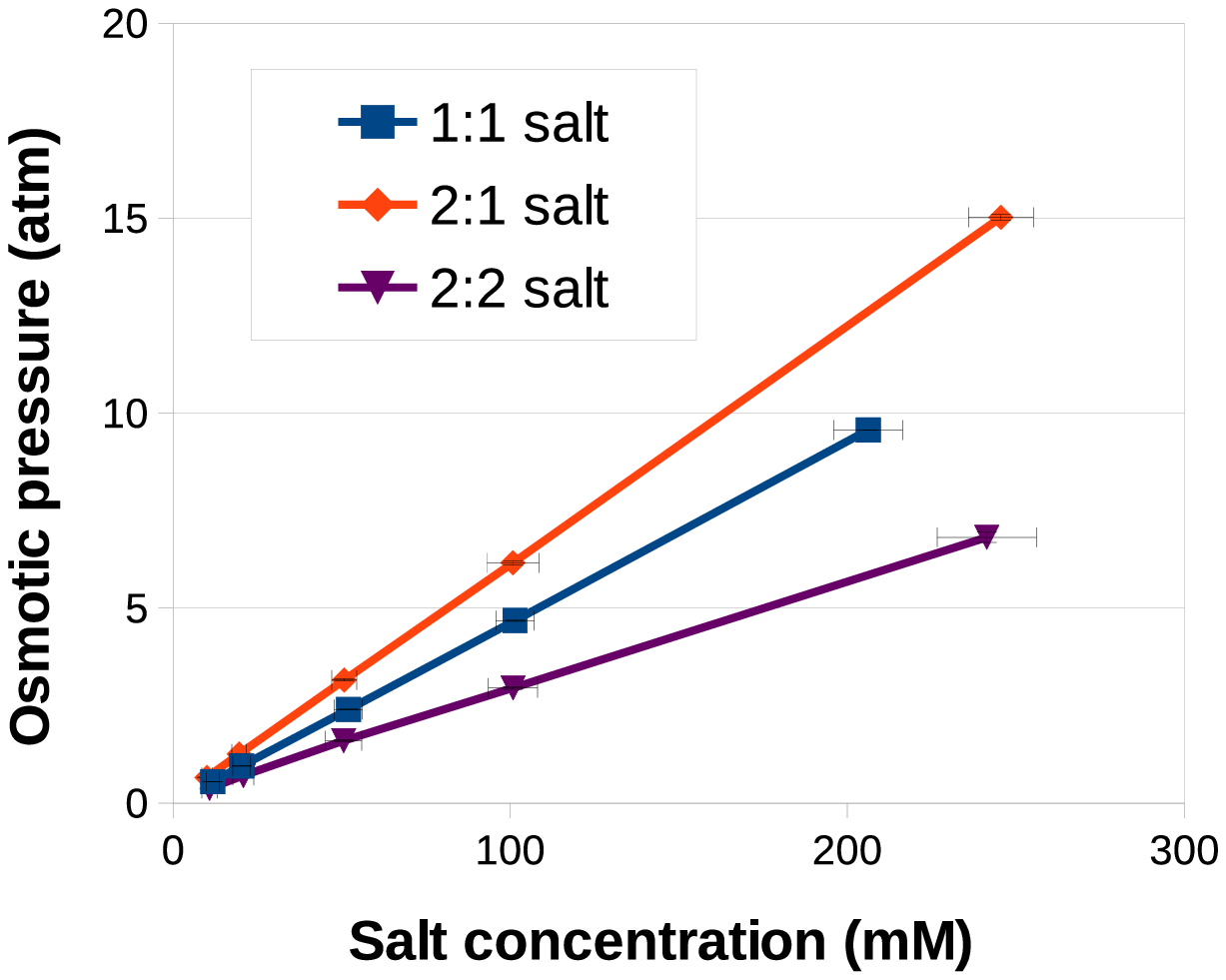}}
\caption{The osmotic pressure of the 
electrolyte solution containing a single type of salt.
The pressure increases linearly with concentration
within the range studied.
}
\label{fig:Psingle}
\end{figure}

Additionally, the omotic pressure of the solution obtained from
simulation is presented in column 3. These values are also plotted
in Fig. \ref{fig:Psingle} for easier comparison. 
\begin{table}
\centering
\begin{tabular}{c||c|c}
\hline
$B_{1:1}/V^2$ (\AA$^{-2}$) & $c$ (mM) & $P_b$ (atm)\\
\hline
$4.00\times 10^{-11}$ & $11.7\pm 1.9$ & $0.552\pm 0.003$\\
$1.15\times 10^{-10}$ & $20.3\pm 2.6$ & $0.954\pm 0.007$\\
$6.60\times 10^{-10}$ & $51.99\pm 4.2$& $2.40\pm 0.012$\\
$2.30\times 10^{-9}$ & $101.4\pm 5.7$ & $4.683\pm 0.023$\\
$8.80\times 10^{-9}$ & $206.2\pm 10.2$ & $9.572\pm 0.001$\\
\hline
\end{tabular}
\caption{The scaled fugacity, $B_{1:1}$ of the 1:1 salt
at different concentrations.
Columns 2 and 3 show the corresponding salt 
concentration and osmotic pressure of the salt bulk solution
obtained from simulation.}
\label{table:mu_single11}
\end{table}
\begin{table}
\centering
\begin{tabular}{c||c|c}
\hline
$B_{2:1}/V^3$ (\AA$^{-3}$) & $c$ (mM) & $P_b$ (atm)\\
\hline
$3.22\times 10^{-16}$ & $10.03\pm 1.56$  & $0.066\pm 0.005$\\
$1.80\times 10^{-15}$ & $19.60\pm 2.19$  & $1.26\pm 0.008$\\
$1.90\times 10^{-14}$ & $50.75\pm 3.69$  & $3.16\pm 0.03$\\
$1.00\times 10^{-13}$ & $100.80\pm 7.71$ & $6.16\pm 0.05$\\
$8.90\times 10^{-13}$ & $245.57\pm 9.63$ & $15.03\pm 0.07$\\
\hline
\end{tabular}
\caption{The scaled fugacity, $B_{2:1}$ of the 2:1 salt for different concentrations.
Columns 2 and 3 show the corresponding salt 
concentration and osmotic pressure of the bulk salt solution
obtained from simulation.}
\label{table:mu_single21}
\end{table}
\begin{table}
\centering
\begin{tabular}{c||c|c}
\hline
$B_{2:2}/V^2$ (\AA$^{-2}$) & $c$ (mM) & $P_b$ (atm)\\
\hline
$6.36\times 10^{-12}$ & $10.03\pm 2.26$ & $0.379 \pm 0.003 $\\
$1.50\times 10^{-11}$ & $20.81\pm 3.07$ & $0.709 \pm 0.028 $\\
$4.45\times 10^{-11}$ & $50.56\pm 5.37$& $1.60\pm 0.016$\\
$9.70\times 10^{-11}$ & $100.81\pm 7.29$ & $2.96\pm 0.033$\\
$2.50\times 10^{-10}$ & $241.39\pm 14.68$ & $6.82\pm 0.130$\\
\hline
\end{tabular}
\caption{The scaled fugacity, $B_{2:2}$ of the 2:2 salt
for different salt concentrations. 
Columns 2 and 3 show the corresponding salt 
concentration and osmotic pressure of the bulk salt solution
obtained from simulation.}
\label{table:mu_single22}
\end{table}
As one can see, at the same concentration, the osmotic pressure of 2:2 salt
solution is lowest, while that of 2:1 salt is highest. This behaviour can be understood. 
Figure \ref{fig:Psingle} shows that, for the concentration
range studied, the osmotic pressure increases linearly with concentration. 
At these low concentrations, our solution should follow the
van der Waals equation of state\cite{landau2013statistical}:
\beq
\left(P+\frac{n^2a}{V^2}\right)(V-nb) = nRT,
\label{eq:vdW}
\eeq
where $n$ is the number of moles of the particles
and $a$, and $b$ are the pressure and volume corrections
due to non-ideality.
The volume correction parameter, $b$, of this equation is small for our system. 
However, the pressure correction parameter, $a$, of the van der Waals
equation of state depends on interactions
among different ions. This is why, at the same concentration,
both 1:1 salt and 2:2 salt contain the same number of ions but
the pressure of 2:2 salt solution is lower due to much stronger attraction
among cations and anions.
On the other hand, for 2:1 salt, there are 3 ions dissolved per molecule compared
to 2 ions dissolved for the other two salts. As a result, the number of moles
of particles are 1.5 times higher than other solution, $n_{2:1} = 1.5 n_{1:1}$, 
leading to higher pressure.

\subsection{Mixture of two salts}

Next steps, we demonstrate the application of our
general GCMC method to simulate the system of two salts, 1:1 salt and
2:1 salt. In a typical physiological experimental setup
with 2:1 divallent salt, one usually has a buffer
solution containing 50mM monovalent salt 
to maintain pH of the solution.
Therefore, in the simulation, we simulate a solution mixture
of 50mM 1:1 salt with varying 2:1 salt concentration.
\begin{table}
\centering
\begin{tabular}{c|c||c|c|c}
\hline
$B_{2:1}/V^{3}$ (\AA$^{-3})$ & $B_{1:1}/V^{2}$ (\AA$^{-2})$ 
	& $c_{2:1}$ (mM) & $c_{1:1}$ (mM) & $P_b$ (atm)\\
\hline
$4.00\times 10^{-15}$ & $8.71\times 10^{-10}$ 
	& $13.9\pm 3.2$ & $50.0\pm 5.9$ & $3.18\pm 0.001$\\
$1.38\times 10^{-14}$ & $1.15\times 10^{-9}$ 
	& $29.9\pm 3.5$ & $50.2\pm 5.0$ & $4.17\pm 0.001$\\
$7.78\times 10^{-14}$ & $1.86\times 10^{-9}$ 
	& $74.7\pm 5.1$ & $50.2\pm 5.3$ & $6.87\pm 0.003$\\
$1.42\times 10^{-13}$ & $2.25\times 10^{-9}$ 
	& $99.8\pm 5.7$ & $50.3\pm 5.4$ & $8.39\pm 0.006$\\
$3.45\times 10^{-13}$ & $3.03\times 10^{-9}$ 
	& $150.2\pm 8.4$ & $50.6\pm 6.7$ & $11.4\pm 0.02$\\
$1.74\times 10^{-12}$ & $5.29\times 10^{-9}$ 
	& $299.6\pm 11.2$ & $49.4\pm 6.8$ & $20.8\pm 0.04$\\
\hline
\end{tabular}
\caption{The scaled fugacities $B_{2:1}$ and $B_{1:1}$ of 1:1 and 2:1 two-salt mixture.
Columns 3 and 4 show the corresponding salt 
concentrations of the simulated bulk solution,
the 1:1 salt concentration is kept approximately constant at 50mM.
Column 5 shows the osmotic pressure of the bulk solution.}
\label{table:mu2}
\end{table}
In columns 3 and 4 of table \ref{table:mu2}, 
the resultant salt concentrations, $c_{2:1}$ and
$c_{1:1}$, of the bulk solution obtained from simulations are listed.
The divalent salt concentration is varied
from 10 mM to 300 mM while the monovalent salt concentration is
kept at approximately 50 mM. 
Note that, even though $c_{1:1}$ is kept constant, 
$B_{1:1}$, (and correspondingly the monovalent salt
chemical potential $\mu_{1:1}$) is not a constant but 
actually increases with $c_{2:1}$, being smallest at $c_{2:1}=0$. 
This is expected because in the system simulated here
we use the same type of monovalent cation for both salts.
Increasing 2:1 salt concentration leads to an
increase in the accompanied concentration of the $-1$ cations.
As a result, the chemical potential of 1:1 salt 
(which is the sum of the chemical potentials of both $+1$ anions and $-1$ cations)
increases.  The linear dependence
of the fugacity $B_{1:1}$ on $c_{2:1}$ shown
in Fig. \ref{fig:B11} agrees with this argument.

\subsection{Mixture of three different salts}
\begin{table}
\centering
\begin{tabular}{c|c|c||c|c|c|c}
\hline
$B_{2:2}/V^{2}$ (\AA$^{-2})$ 
	& $B_{2:1}/V^{3}$ (\AA$^{-3})$ & $B_{1:1}/V^{2}$ (\AA$^{-2})$ 
	& $c_{2:2}$ (mM) & $c_{2:1}$ (mM) & $c_{1:1}$ (mM) & $P_b$ (atm)\\
\hline
$5.86\times 10^{-12}$ & $3.86\times 10^{-15}$ & $7.58\times 10^{-10}$ 
	& $9.9\pm 3.0$ & $10.1\pm 3.9$ & $49.9\pm 6.2$ & $3.26\pm 0.04$\\
$1.36\times 10^{-11}$ & $4.70\times 10^{-15}$ & $7.15\times 10^{-10}$ 
	& $20.2\pm 4.3$ & $9.8\pm 4.9$ & $49.8\pm 6.4$ & $3.54\pm 0.03$\\
$4.13\times 10^{-11}$ & $6.74\times 10^{-15}$ & $6.63\times 10^{-10}$ 
	& $49.9\pm 6.7$ & $10.1\pm 7.0$ & $49.9\pm 6.5$ & $4.40\pm 0.01$\\
$9.45\times 10^{-11}$ & $8.75\times 10^{-15}$ & $6.02\times 10^{-10}$ 
	& $99.7\pm 9.3$ & $ 9.9\pm 9.5$ & $50.0\pm 6.7$ & $5.86\pm 0.02$\\
$1.50\times 10^{-10}$ & $1.03\times 10^{-14}$ & $5.71\times 10^{-10}$ 
	& $150.1\pm 11.4$ & $10.2\pm 11.5$ & $50.0\pm 6.8$ & $7.25\pm 0.03$\\
$2.05\times 10^{-10}$ & $1.14\times 10^{-14}$ & $5.50\times 10^{-10}$ 
	& $200.0\pm 15.5$ & $10.1\pm 15.5$ & $50.2\pm 6.8$ & $8.33\pm 0.30$\\
$3.30\times 10^{-10}$ & $1.35\times 10^{-14}$ & $5.22\times 10^{-10}$ 
	& $307.7\pm 15.0$ & $9.9\pm 19.4$ & $50.1\pm 6.8$ & $11.4\pm 0.05$\\
\hline
\end{tabular}
\caption{The scaled fugacities, $B_{2:2}$, $B_{2:1}$ 
and $B_{1:1}$, of the three salts in the mixture. Columns 4, 5 and 6 show the corresponding salt 
concentrations. The 1:1 salt concentration is kept approximately constant at 50mM. The 2:1 salt concentration is kept approximately constant at 10mM. Column 7  shows the osmotic pressure of the bulk solution obtained from simulation.
}
\label{table:mu3}
\end{table}
\begin{figure}
\centering
\begin{minipage}{.5\textwidth}
  \centering
  \resizebox{7.2cm}{!}{\includegraphics{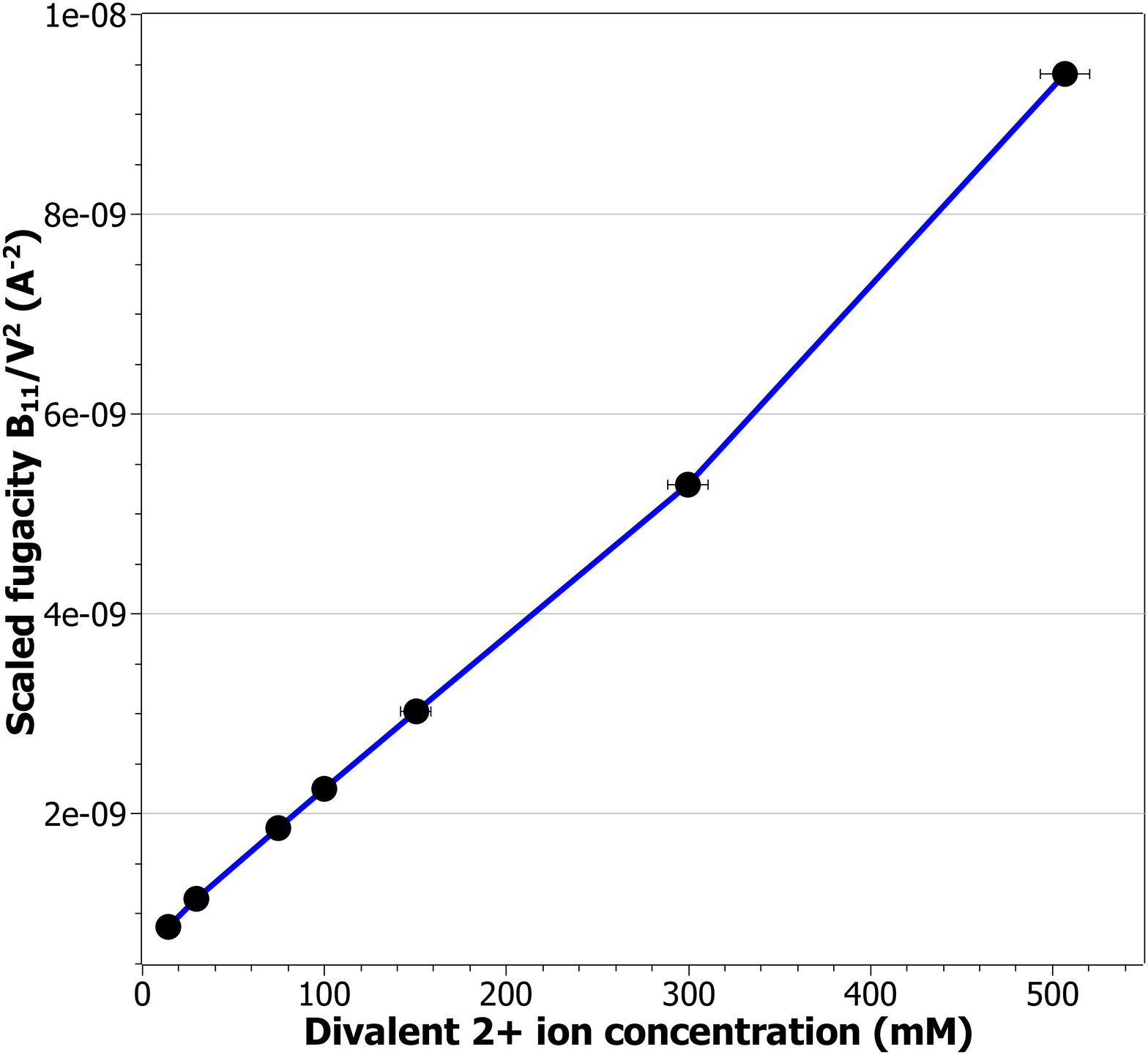}}\\
  a)
\end{minipage}%
\begin{minipage}{.5\textwidth}
  \centering
  \resizebox{7.5cm}{!}{\includegraphics{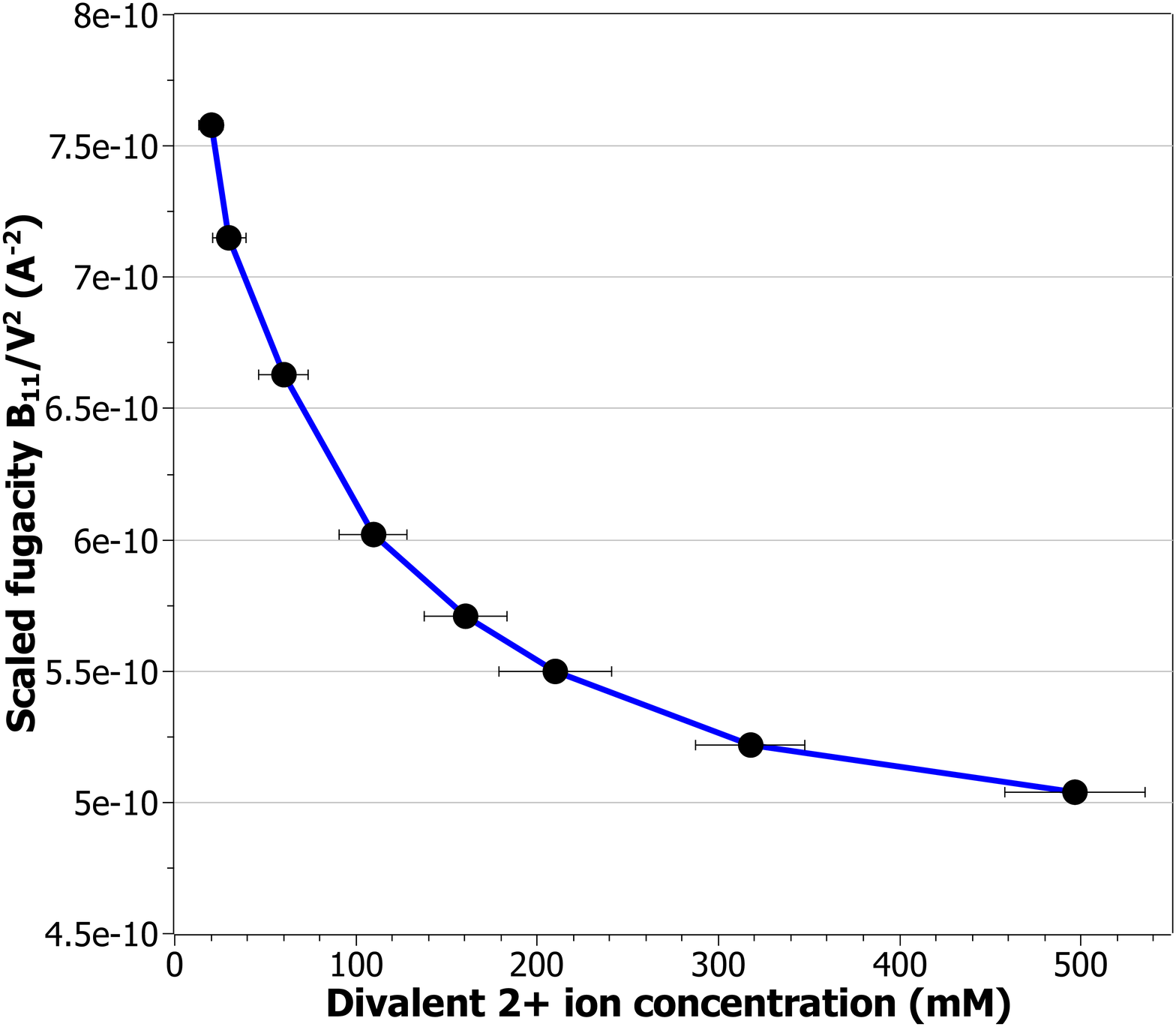}}\\
  b)
\end{minipage}
\caption{The scaled fugacity $B_{11}$ of 1:1 monovalent salt
as function of the divalent anion concentrations. It increases
with increasing divalent anion concentrations for
the case of no divalent cations (a); it decreases with
increasing divalent anions in the presence of divalent
cations (b). In both case, the simulated solution contains about 50mM 
monovalent salt.
The opposite behaviors of $B_{11}$ shows that divalent cations
make it easier to insert monovalent cations into solution
(see text for more discussion).}
\label{fig:B11}
\end{figure}
Lastly, we report on the more complicated but experimentally relevant case of an electrolyte solution where all three 2:2, 2:1 and 1:1 salts are present. Such a mixture is used in
studying the effect of MgSO$_4$ salt on DNA ejection from bacteriophages
such as that done in Ref. \cite{Knobler08}. While the MgSO$_4$ salt (2:2 salt) concentration varies,
the pH of the solution is maintained using a buffer solution of 10mM MgCl$_2$ and 50mM NaCl. Consequently, in this section, we report on simulation of solution containing varying concentration of 2:2 salt, while keeping the 1:1 salt concentration fixed at 50mM, and 2:1 salt
concentration fixed at 10mM. The monovalent cations are
assumed to be the same for both salts (Cl$^{-1}$ ions
in experiments).
Table \ref{table:mu3} list the corresponding fugacity parameters $B$ 
of the three salts in solution. Similar to previous
sections, the resultant concentrations of the ion species
are reported in columns 4, 5 and 6. 
The osmotic pressure of the solution is reported
in the last columns. A very important physical observation 
for this three salt mixture is the fact that, in the presence of
divalent cations, the fugacity of monovalent salt {\em decreases}
with {\em increasing} divalent anion concentrations. In other
words, divalent cations make it easier to insert
monovalent cations into the solution. This is show clearly
in Fig. \ref{fig:B11} where the scale fugacity $B_{11}$ 
is plotted for different divalent anion concentrations
without (Fig. \ref{fig:B11}a) and with (Fig. \ref{fig:B11}b)
the presence of divalent cations. 

The opposite behaviors of $B_{11}$ 
shows that divalent cations make it easier to insert monovalent 
cations into solution. Physically, one could explain this
behaviour by the fact that it takes one divalent cations instead
of two monovalent cations to conpensate for the charges
of divalent anions. Entropically, this leaves ``room" in the system
to add more monovalent salt to the solution. 
Experimentally, there is quantitative difference between DNA ejection 
from bacteriophages in MgCl$_2$+NaCl solution and MgSO$_4$+MgCl$_2$+NaCl 
solution \cite{Knobler08}. Our simulation showing monovalent salt
can enter bacteriophages easier in the presence of MgSO$_4$ (2:2)
salt qualitatively agrees with the experimental fact that
that MgSO$_4$ make DNA ejection easier. Nevertheless, more
detail investigation is needed. This will be studied in a future work.

Before conluding this section, let us show some preliminary result on
the effect of ion sizes on the fugacities and the pressure of 
electrolyte solution. Mixture of 2:1 and 1:1 salts is studied
with the 1:1 salt concentration approximately 50mM. The 
2:1 salt concentration is varied from about 20mM to 500mM. The
radius of the divalent cation is simulated at 2\AA, 2.5\AA and
3\AA\ respectively. 

\begin{figure}
\centering
\begin{minipage}{.5\textwidth}
  \centering
  \resizebox{7.2cm}{!}{\includegraphics{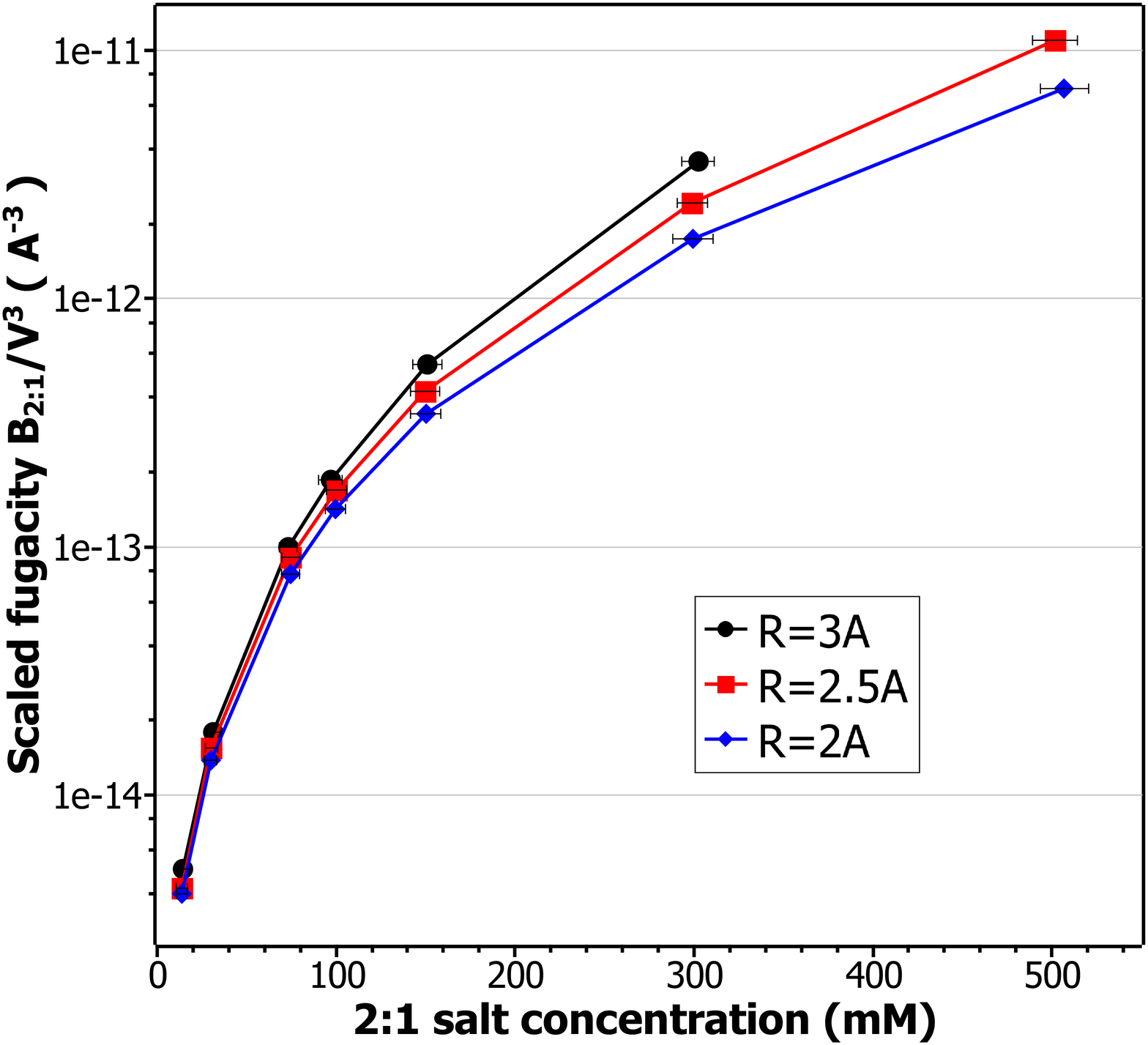}}\\
  a)
\end{minipage}%
\begin{minipage}{.5\textwidth}
  \centering
  \resizebox{7.5cm}{!}{\includegraphics{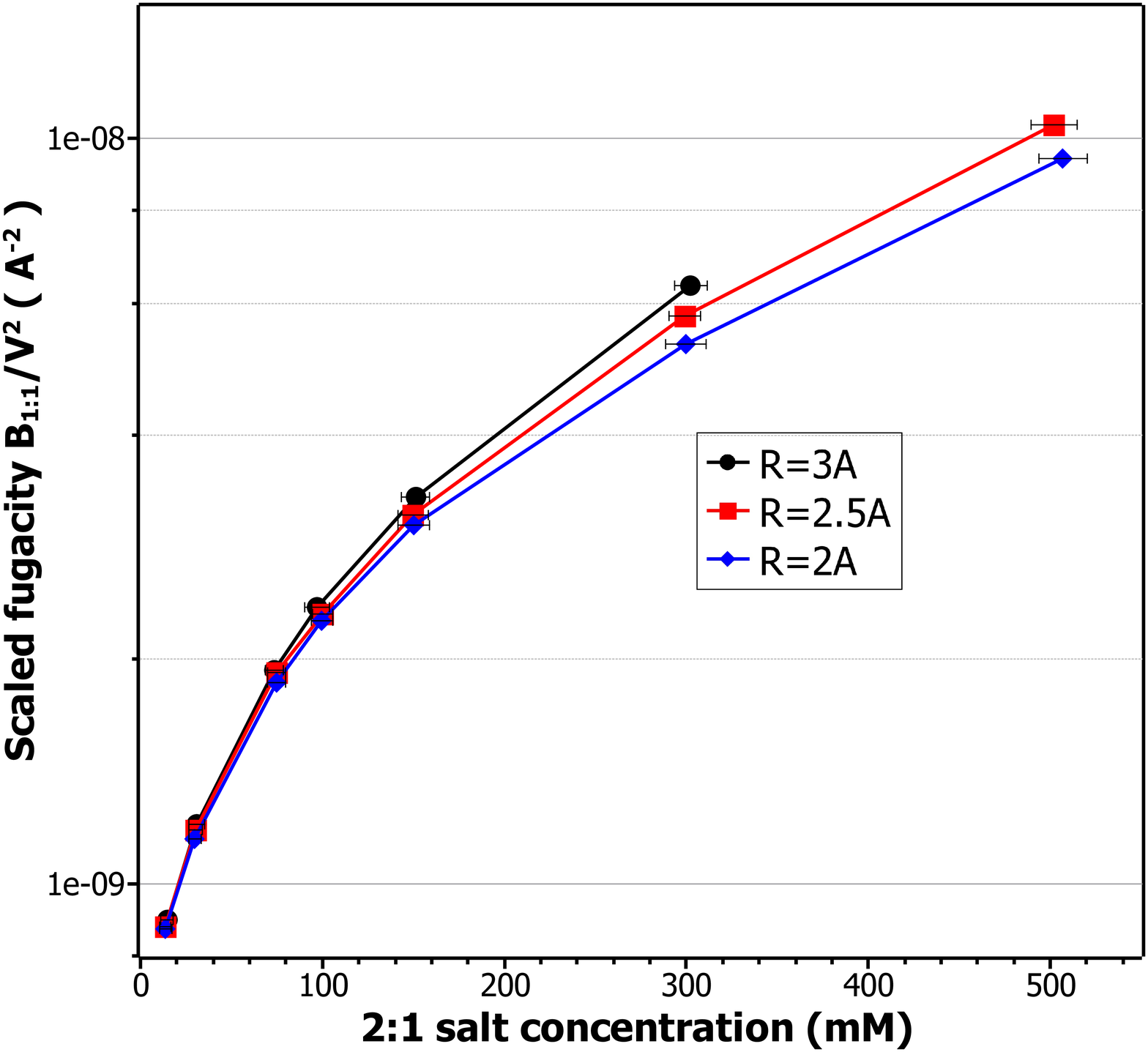}}\\
  b)
\end{minipage}
\begin{minipage}{.5\textwidth}
  \centering
  \resizebox{7.5cm}{!}{\includegraphics{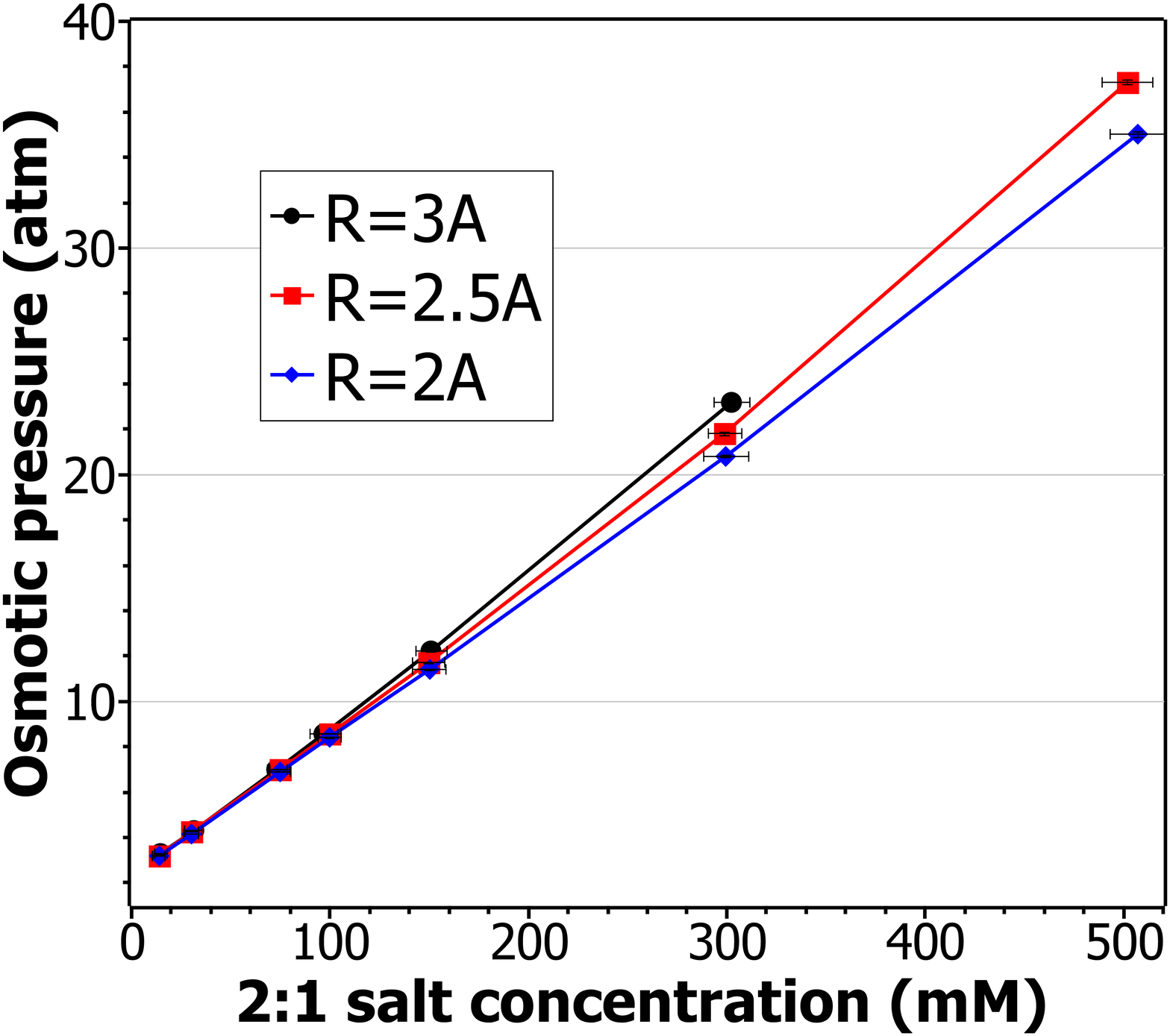}}\\
  c)
\end{minipage}
\caption{The scaled fugacity $B_{11}$ of 1:1 monovalent salt (a)
and $B_{21}$ of 2:1 divalent salt (b) and the osmotic pressure
(c) in a mixture of the two salts as function of the divalent anion concentrations. 
Different
curves corresponding to different value of the radius
of the divalent anion, 2\AA, 2.5\AA, and 3\AA. 
(see text for more discussion).}
\label{fig:Posm_R}
\end{figure}
In Fig. \ref{fig:Posm_R}, the fugacities of 2:1 salt and 1:1 salt in the mixture
and the osmotic pressure is plotted. They show that increasing the size
of $+2$ anions leads to increased fugacities and pressure at higher concentrations.
This is inline with the van der Waals equation of state where
the volume correction terms $b$ becomes larger for larger
ions. Nevertheless, within the size range studied, this leads to about
less than 10\% corrections to the fugacity and pressure up to 500mM divalent salt
concentration.

\section{Conclusion\label{sec:conclusion}}

In this paper, we presented an extensive study of the Grand-Canonical 
Monte-Carlo simulation for electrolyte solutions using a primitive ion mode.
Application of this method to simulate  various solutions
containing single salt, two salt mixture and three salt mixture are 
carried out. The fugacities of individual salt species for different
solutions at typical concentrations are reported. The result
of osmotic pressure of the electrolyte solution are calculated
and shown to be linearly proportional to the salt concentration
within the range of concentrations considered. However, the
pressure differs for different type of salt because the non-ideal
gas corrections are different for different ion valence.

Comparing the solution without and with divalent cations, it
is shown that divalent cations make it easier to insert 
monovalent cations in solution. This agrees qualitatively
with experimental result of DNA ejection from bacteriophages
in MgSO$_4$ salt mixtures and MgCl$_2$ salt mixtures.

In this paper, the aqueous  solution is simulated implicitly. It appears only in the dielectric constant of the medium. Our method is suitable therefore for a coarse-grained region in a multiscale simulation setup. If one simulates the solvent molecules explicitly, it is likely that a full particle insertion or deletion would be impractical due to a large change in the system energy. In such case, partial deletion/insertion of particle is preferable. Nevertheless, it is very unlikely one would practically need grand-canonical simulation in the atomistic region in a multiscale simulation.

\section{Acknowledgments}

We would like to thank Drs. T. X. Hoang and Paolo Carloni for valuable discussions. 
TTN acknowledges the financial support of
the Vietnam National University grant number QG.16.01 and, partially by the USA National Science Foundation grant NFS CBET-1134398.
The authors are indebted to Dr
A. Lyubartsev for providing us with the 
Fortran source code of their Expanded Ensemble Method.


\bibliography{nttpaper}

\end{document}